\documentclass[aps,superscriptaddress,nofootinbib,twocolumn]{revtex4-1}

\usepackage{mathrsfs}
\usepackage{amsfonts}
\usepackage{amssymb}
\usepackage{amsmath}
\usepackage{graphicx}
\usepackage[usenames,dvipsnames]{color}
\usepackage[colorlinks=true,citecolor=blue,linkcolor=magenta]{hyperref}
\usepackage{ulem}
\usepackage{lmodern}

\newcommand{\figpath}{.}

\newcommand{\Tr}{\mathrm{Tr}}

\newcommand{\ket}[1]{\vert{ #1 }\rangle}

\newcommand{\SIGL}[2]{\sigma^{\mathrm{#1}}_{#2}}
\newcommand{\SIGP}[2]{\tilde{\sigma}^{\mathrm{#1}}_{#2}}
\newcommand{\TP}[2]{T^{\mathrm{#1}}_{#2}}
\newcommand{\Ide}{\openone}

\begin{document}
\title{Stabilisers as a design tool for new forms of Lechner-Hauke-Zoller Annealer}
\author{Andrea Rocchetto, Simon C. Benjamin, Ying Li}
\affiliation{Department of Materials, University of Oxford, Parks Road, Oxford OX1 3PH, United Kingdom}
\begin{abstract}
In a recent paper Lechner, Hauke and Zoller (LHZ) described a means to translate a Hamiltonian of $N$ spin-$\frac{1}{2}$ particles with `all-to-all' interactions into a larger physical lattice with only on-site energies and local parity constraints. LHZ  used this mapping to propose a novel form of quantum annealing. Here we provide a stabiliser-based formulation within which we can describe both this prior approach and a wide variety of variants. Examples include a triangular array supporting all-to-all connectivity, and moreover arrangements requiring only $2N$ or $N\log N$ spins but providing interesting bespoke connectivities. Further examples show that arbitrarily high order logical terms can be efficiently realised, even in a strictly 2D layout. Our stabilisers can correspond to either even-parity constraints, as in the LHZ proposal, or as odd-parity constraints. Considering the latter option applied to the original LHZ layout, we note it may simplify the physical realisation since the required ancillas are only spin-$\frac{1}{2}$ systems (i.e.~qubits, rather than qutrits) and moreover the interactions are very simple. We make a preliminary assessment of the impact of this design choices by simulating small (few-qubit) systems; we find some indications that the new variant may maintain a larger minimum energy gap during the annealing process.
\end{abstract}

\maketitle

\section{Introduction}

Quantum annealing (QA) is an an approach to solving optimisation problems, a family of tasks that include many important examples ranging from financial portfolio management to vehicle routing~\cite{kadowaki1998quantum, farhi2001quantum}. Typically the task can be thought of as minimising a {\it cost function} that depends upon many variables. In QA this is done by considering a physical system whose energy corresponds to the cost, and seeking that system's ground state. One can implement the QA approach using conventional hardware by running an algorithm that mimics quantum behaviour~\cite{martovnak2002quantum}; this is then a variant of classical simulated annealing~\cite{kirkpatrick1983optimization}. Alternatively one can aim to construct real quantum technologies whose components are indeed quantum entities capable of superposition and entanglement. The annealing process can exploit the adiabatic theorem in order to remain in, or near, the system's ground state when changing parameters (such as a global field) sufficiently slowly. By starting from a Hamiltonian whose ground state can be reliably achieved, and morphing slowly to a final Hamiltonian whose energies reflect the cost in the optimisation problem, the hope is that measuring the final state reveals a low cost solution.

There are many interesting questions associated with this approach. The prospects of reaching the ground state of the final Hamiltonian will depend on rate at which the Hamiltonian is changed versus the size of the {\it gap} from ground state to lowest excited states; the smaller the gap, the slower the evolution must be \cite{childs2001robustness}. However since the gap size cannot be pre-computed for problems of meaningful size, it is difficult to be definite about how fast the system can be permitted to evolve, or indeed whether the approach can succeed at all given finite temperatures. Thus the performance of a QA system is hard to predict analytically. Prototype systems do exist from the company D-Wave, and several studies have sought to evaluate the power of these systems by empirical testing (see e.g. Ref.~\cite{ronnow2014defining}). One can also make comparisons with QA simulated by quantum Monte Carlo, although this must be done with caution as there are subtleties with the discretisation of time \cite{heim2015quantum}.

Another important question is that of connectivity. In contrast to conventional computers, and indeed circuit model quantum computers, the adiabatic approach involves keeping interactions between qubits ``always-on'' so as to maintain the energy gap. This implies that the qubits, or {\it spins} as we will henceforth refer to them, should maintain direct physical interactions continuously. Whereas ideally one might wish to be able to connect any physical spin in the device to any other, in practice this is inconsistent with interactions which are implemented through short range physical links. In the D-Wave chips, the set of permitted non-zero links between the physical spins is called Chimera graph (see Ref.~\cite{bunyk2014architectural}). It is locally rich, but on the large scale it has the form of a two dimensional nearest-neighbour lattice. Typically for a real-world optimisation problem, such as the satisfiability problems \cite{jorg2010energy, jorg2010first}, one would not expect that the variables are tensioned against each other in a pattern that respects any particular geometry. Therefore for the {\it logical problem} to be realised as a physical annealing task, it must in some way be re-expressed.

One solution is based on minor embedding \cite{choi2008minor, choi2011minor, boothby2016fast, cai2014practical}: in effect groups of physical spins are bound together with very strong interactions in order to form extended single spin entities. These larger entities have correspondingly more connections to one another. In order to achieve all-to-all connectivity in this way, the $N$ spins of the logical problem must be encoded into Order($N^2$) physical spins. However, even assuming that this cost is permissible (and one should expect that the increased number of physical spins corresponds to a reduction in the crucial energy gap), there is a question as to whether this approach is practical. When a large number of physical spins are bound together with achievable interaction strengths, it is not clear that the extended objects will indeed function as equivalent to single logical spins.

An alternative formulation of the mapping was recently proposed by Lechner, Hauke and Zoller (LHZ) in Ref.~\cite{Lechner2015}. According to this approach, the physical spins now represent the links, or relative orientations, between the logical spins. Thus there is one {\it physical} spin whose role is to represent the relative orientation of {\it logical} spins 1 and 2: if they are aligned ($\uparrow\uparrow$ or $\downarrow\downarrow$) then the physical spin will take one value ($\uparrow$, say), whereas if the logical spins are anti-aligned  ($\downarrow\uparrow$ or $\uparrow\downarrow$) then the physical spin takes the opposite orientation ($\downarrow$). Since there are $N(N-1)/2$ possible pairings of the $N$ logical spins this leads to the same Order($N^2$) resource cost as the minor embedding approach; however one avoids the need to bind multiple physical spins into single entities, and indeed the coupling strengths in the logical model conveniently map to single-spin energies in the physical hardware. {\it Note added in late draft: a paper by Albash, Vinci, and Lidar has very recently been posted, which makes a detailed comparison of minor embedding versus the LHZ approach, see Ref.~\cite{lidar2016}.}

Here we view the LHZ approach within the general framework of a set of stabilisers, which allows us to recover the LHZ construction as well as a wide variety of variations. Each stabiliser is a product of physical $z$ operators which we constrain to one of its eigenvalues, either $+1$ or $-1$. This can be viewed as a restriction to either even, or odd, parity. The original LHZ construction corresponds to all-even stabilisers; we find there there may be some advantages to switching to all-odd constraints instead; the required ancilla structure is more simple and moreover our simple numerical simulations indicate that level crossings are less frequent. 

In addition to the square lattice of the LHZ proposal, our stabiliser formulation leads us to a number of interesting alternatives. If all-to-all connectivity is required, then triangular (three-body) stabilisers can replace the four-body stabilisers of LHZ. Moreover if a lower level of connectivity will suffice then our approach can provide layouts involving fewer physical spins. Note that some NP problems, like graph colouring, are hard to solve (or indeed, {\it hardest} to solve) when they don't have a fully connected graph~\cite{cheeseman1991really}. We give examples where $2N-1$ and $N\log N$ physical spins realise non-trivial connectivities between $N$ logical spins. Additionally, we show how arbitrarily high order terms in the logical Hamiltonian (such as $\sigma^Z_i\sigma^Z_j\dots\sigma^Z_m$) can be mapped to a single spin in the physical layout.

Our approach is conceptually straightforward. We take a candidate layout of $N_P$ physical spins, and we specify $N_S=N_P-N$ stabiliser constraints. We then nominate $N$ of the physical spins, each of which will correspond to a logical spins in the following sense: the $z$-operator a physical spin is identified with the same operator on the logical spin. Finally we identify the logical $x$-operators that are implied by these earlier choices; each will be a product of operators forming a chain that crosses the layout, rather analogously to logical operators in topological codes such as the Kitaev's surface code \cite{bravyi1998quantum}. Intersections between these logical $x$ chains allow us to find the meaning of each individual physical spin, i.e. to identify what product of logical $z$ spins it represents.  

\section{Parity-constraint annealing and stabiliser code}

We define the {\it logical} spin glass model, in which each spin can have a non-zero interaction with every other as well as an arbitrary local field, as follows
\begin{eqnarray}
H_\mathrm{logic} = \sum_{i = 1}^{N} h_{i} \SIGL{Z}{i} + \sum_{i = 1}^{N-1} \sum_{j = i+1}^{N} J_{i,j} \SIGL{Z}{i}\SIGL{Z}{j}.
\label{logicalH}
\end{eqnarray}
Note this Hamiltonian is general in the sense that the local fields $h_{i}$ and the interaction strengths $J_{i,j}$ can take any value. However it does not contain three-body or higher interaction terms, which would be convenient for optimising functions containing terms with three or more variables involved. In fact, in the final part of the analysis presented here we will extend our considerations to logical Hamiltonians that {\it do} contain arbitrary higher order terms. For simplicity we will now focus on the case where the logical Hamiltonian has the form above.

\begin{figure*}
\centering
\includegraphics[scale=.7]{\figpath 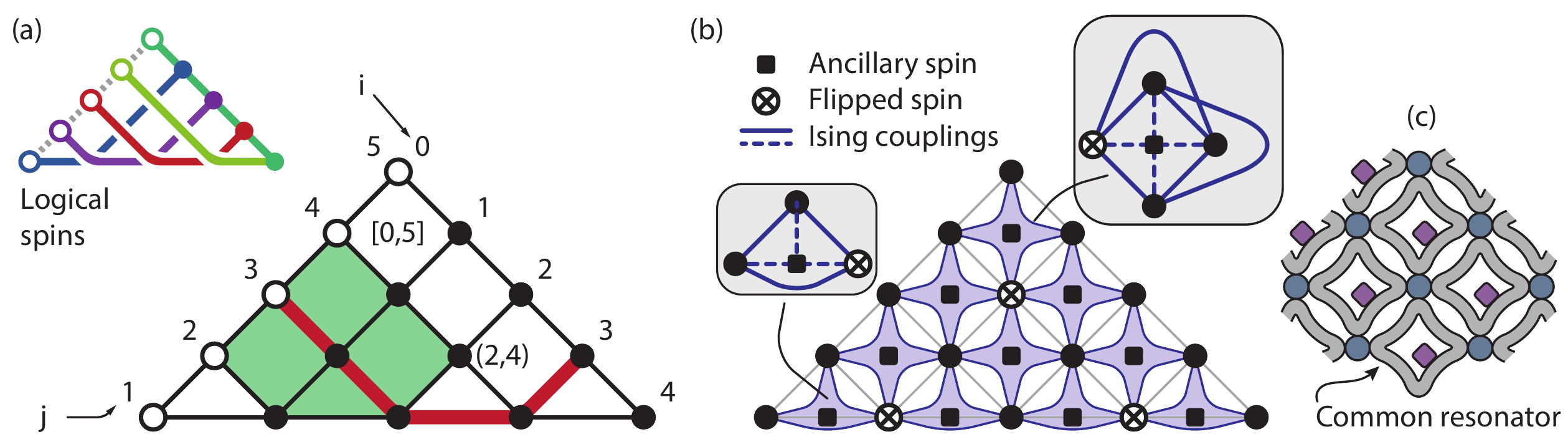}
\caption{
The lattice of $N(N+1)/2$ physical spins for encoding $N$ logical spins. (a) The formal scheme, (b) a physical implementation using ancilla qubits, and (c) a schematic indicating that common resonators might mediate the interactions.
}
\label{fig:squareLattice}
\end{figure*}

The task now is to successfully emulate the physics of this ideal, logical Hamiltonian using a real architecture in which a larger number of physical spins interact only locally. We begin by selecting a layout for the physical spins; our first choice will be the two-dimensional lattice with a square unit cell as proposed by LHZ (see Fig.~\ref{fig:squareLattice}). This structure contains $N(N+1)/2$ physical spins, therefore its full Hilbert space is vastly greater than that of the logical Hamiltonian: we must apply constraints to define a suitable subspace. We will specify a set of mutually-commuting {\it stabilisers}, each being a operator formed by a product of single-spin Pauli operators. We will require that the state of the system is a mutual eigenstate of all these stabilisers with a specified eigenvalue in each case: either $+1$ or $-1$. (Note this is a slight departure from the usual convention where all stabiliser eigenvalues are $+1$ because any negative value is absorbed into the definition of the stabiliser itself). As we enforce each such stabiliser, we will halve the dimension of the compatible Hilbert space. Therefore we will require $N(N+1)/2-N=N(N-1)/2$ stabilisers in order that the compliant subspace has the desired dimension $2^N$. Our remaining task will be to identify observables in the physical lattice that correspond to the measurement of single spins in the logical Hamiltonian, and establish the conditions under which the correspondence is correct.

As noted, the physical layout is a lattice of square cells forming a triangular shape as shown in Fig.~\ref{fig:squareLattice}. There are $N(N+1)/2$ spins. Presently, we will introduce additional ancilla spins whose role is simply to enforce the stabiliser constraints in a physically natural way -- but when we allude to {\it physical spins} without explicitly using the term {\it ancilla} then we are referring to the $N(N+1)/2$ spins that form the direct physical embodiment. We label each physical spin with two coordinates $(i,j)$ and each face according to the spin at the top corner, e.g.~the top face is labelled as $[0,5]$ in Fig.~\ref{fig:squareLattice}. The reason for this labelling will become apparent; for now is it simply a systematic way to uniquely tag each spin.

In order that this formulation is consistent with the LHZ approach, we select a family of stabilisers that can constrain the parity of the physical spins around each face of the lattice. Face $[i,j]$ corresponds to a stabiliser
\begin{eqnarray}
S_{[i,j]} = \left\{
\begin{array}{ll}
\SIGP{Z}{(i,j)}\SIGP{Z}{(i,j-1)}\SIGP{Z}{(i+1,j)}\SIGP{Z}{(i+1,j-1)}, & \mbox{if $i+2 < j$;\ \ \ }\\
\SIGP{Z}{(i,j)}\SIGP{Z}{(i,j-1)}\SIGP{Z}{(i+1,j)}, & \mbox{if $i+2 = j$.\ \ \ \ }
\end{array} \right.
\end{eqnarray}
where the second case $i+2=j$ simply corresponds to the triangular faces along the base of the lattice. The tilde symbol is used to indicate that the sigma operators $\SIGP{}{}$ apply to physical spins.
Each stabiliser is thus the product of Pauli operators acting on the physical spins around the face. Since each Pauli operator has eigenvalues $\pm 1$ corresponding to its spin being orientated in the {positive/negative} $z$-direction, it follows that the stabiliser's eigenvalues are $\pm 1$. This value can be interpreted as a parity if we take $+1$ to indicate `even' and $-1$ to indicate `odd' in terms of the the number of spins aligned along negative $z$-direction. This odd/even label is natural for four-body stabilisers, but can be less intuitive for other stabilisers -- therefore we will use the term odd/even only for the four-body case, and more generally we will speak of positive and negative stabilisers. 

We will require that the states of interest $\ket{L}$ are good eigenstates of {\it all} these stabilisers, i.e.
\[
S_{[i,j]} \ket{L}=\nu_{i,j}\ket{L},
\]
where we will specify each individual $\nu_{i,j}$ as $+1$ or $-1$. Equivalently, writing $P_0$ as the projector into the legitimate subspace, we require $S_{[i,j]} P_0=\nu_{i,j}P_0$. We easily verify  that this set of stabilisers mutually commute (all are $z$-basis operators) and moreover they are independent: specifying the eigenvalues of any subset does not constrain the eigenvalues of the remaining ones. This latter property is confirmed by noting that as we consider each new face of the lattice, we are encountering at least one new qubit. (Note that in contrast, the 2D toric stabiliser code has the feature that specifying all-but-one of the stabilisers in a given basis will logically imply the value of the last one). It is worth remarking that the legitimate states of the physical spins are formally {\it stabiliser states} if we opt to absorb the $\nu_{i,j}$ factors into the stabilisers; but we find it convenient to regard the stabilisers as fixed entities and the $\nu_{i,j}$ as their required eigenvalues.

It remains to identify observable properties of the physical lattice which will correspond to the single-spin operators {$\SIGL{Z}{i}$ (and $\SIGL{X}{i}$)} in the logical Hamiltonian. We call these the {\it logical spin operators}. We will simply choose one set, the $\SIGL{Z}{i}$ set, and then attempt to identify the appropriate $\SIGL{X}{i}$. Remaining consistent with the LHZ construction, we assign our logical $Z$ operators to the individual physical spins on the left-side diagonal, i.e.
\begin{eqnarray}
\SIGL{Z}{k} = \SIGP{Z}{(0,k)}.
\label{eq:logical_Z}
\end{eqnarray}
Clearly these logical operators commute both with one another and with the stabilises (all are $z$-basis) and they are  independent from one another and from the stabilisers (specifying the states of these logical observables would not constrain any of the stabiliser eigenvalues). Therefore this choice is valid. With this choice, one can of course read the state of logical spins in the $z$-basis by measuring physical spins on the left side of the lattice.

We must now identify the logical spin $x$ operators with the same properties of commutation and independence, except that of course that the $z$ and $x$ operators for the same logical qubit should anti-commute. No set of operators on single physical spins will have this property; instead we should look for products of physical spin operators. Consider the logical $x$-operator for logical spin $i=3$. Since it must anti-commute with logical operator $ \SIGP{Z}{(0,3)}$ we include $\SIGP{X}{(0,3)}$ in the product of operators. However, it should commute will all our earlier stabilisers, so it should include zero, two or four $\SIGP{X}{(0,3)}$operators around each face of the lattice (i.e. each stabiliser). Finally, it should not include $\SIGP{X}{(0,k)}$ for any $k\neq 3$ or else it will not commute with those other logical spins. Considering these constraints, we are led to the unique solution: a product of $\SIGP{X}{\ }$ operators along the path indicated in red in Fig.~\ref{fig:squareLattice}. The other logical $x$-operators have analogous forms, descending from the left side and `bouncing' from the base as shown in the upper left inset in Fig.~\ref{fig:squareLattice}. Formally,
\begin{eqnarray}
\SIGL{X}{k} = \prod_{i}\SIGP{X}{(i,k)} \prod_{j}\SIGP{X}{(k,j)},
\label{eq:logical_X}
\end{eqnarray}
where $k = 1,2,\ldots,N$. That is, the logical $\SIGL{X}{k}$ operator is the product of Pauli operators of the set of physical spins with a label either $i=k$ or $j=k$.


At this point we are ready to re-write the original logical spin Hamiltonian in terms of operators on the physical lattice. Each term of the form $h_j\SIGL{Z}{j}$ simply translates to $h_j\SIGP{Z}{(0,j)}$. Meanwhile, each term $J_{i,j} \SIGL{Z}{i}\SIGL{Z}{j}$  translates directly to  $J_{i,j} \SIGP{Z}{(0,i)}\SIGP{Z}{(0,j)}$ but crucially this can be rewritten as a single physical spin operator $\mu_{i,j}J_{i,j} \SIGP{Z}{(i,j)}$, where $\mu_{i,j}$ is a certain product of our $\nu_{i,j}=\pm 1$ eigenvalues. To see this, consider first the physical spin at $(1,2)$, and moreover the value of operator $\SIGP{Z}{(1,2)}$. For all legitimate states, $S_{[0,2]}\ket{L}=\nu_{0,2}\ket{L}$. But $S_{[0,2]}$ is defined as $\SIGP{Z}{(0,1)}\SIGP{Z}{(0,2)}\SIGP{Z}{(1,2)}$, and the first two of these operators are logical $Z$ operators and therefore must be free to take any value, i.e. we can make no constraint on  $\SIGP{Z}{(1,0)}\ket{L}$, $\SIGP{Z}{(2,0)}\ket{L}$ or their product. Then, in order to ensure that $S_{[0,2]}\ket{L}=\nu_{0,2}\ket{L}$ we must insist that $\SIGP{Z}{(1,2)}\ket{L}=\nu_{0,2}\SIGP{Z}{(0,1)}\SIGP{Z}{(0,2)}\ket{L}$. Thus the physical spin at $(1,2)$ is entirely dependent on the physical spins at $(0,1)$ and $(0,2)$, in such a way as to satisfy the face stabiliser. Consequently for the legitimate subspace we can rewrite the term $J_{1,2} \SIGP{Z}{(0,1)}\SIGP{Z}{(0,2)}$ as $\nu_{0,2}J_{1,2}\SIGP{Z}{(1,2)}$.

Having thus established the dependent nature of the physical spin at $(1,2)$, one can repeat the argument for the physical spin at $(1,3)$:  the face stabiliser $[0,3]$ involves two logical spins, and the spin whose dependence we have just determined, thus the dependence of the fourth spin is implied. We find that in order to ensure $S_{[0,3]}\ket{L}=\nu_{0,3}\ket{L}$, we will require $\SIGP{Z}{(1,3)}\ket{L}=\nu_{0,2}\nu_{0,3}\SIGP{Z}{(0,1)}\SIGP{Z}{(0,3)}\ket{L}$. One can proceed to establish the dependence of every remaining physical spin; it is always of the form $\SIGP{Z}{(i,j)}\ket{L}=\mu_{i,j}\SIGP{Z}{(0,i)}\SIGP{Z}{(0,j)}\ket{L}$ where $\mu_{i,j}$ is simply a certain product of our chosen stabiliser eigenvalues:
\begin{eqnarray}
\mu_{i,j} = \prod_{i'=0}^{i-1} \prod_{j'=i+1}^{j} \nu_{i',j'}\ \ \ \ {\rm thus}\ \ \ \ \mu_{i,j}=\pm 1.
\label{eqn:muRule}
\end{eqnarray}
In general this is the product of the $\nu$ values in a block of the array; in Fig.~\ref{fig:squareLattice} the green block corresponds to the set of $\nu$ values that must be multiplied to find $\mu_{2,4}$.

Now since we chose to identify $\SIGP{Z}{(0,i)}$ with the logical $z$-operator $\SIGL{Z}{i}$, we can now conclude that each spin $(i,j)$ encodes the logical product $\SIGL{Z}{i}\SIGL{Z}{j}$ (up to the sign $\mu_{i,j}$). Thus the motivation for our labelling scheme is apparent. We note from the Fig.~\ref{fig:squareLattice}(a) inset that physical spin $(i,j)$ also lies at the interaction of the logical $x$-operator chains for logical spins $i$ and $j$. In a later section we will show that this is generally the case, for any lattice: if the logical $x$-operator for logical spin $i$ intersects with a given physical spin, then $\SIGL{Z}{i}$ is in the product of logical $z$-operators to which that spin's physical $z$-operator $\SIGP{Z}{(0,i)}$ corresponds. This is a more efficient way to identify the roles of physical spins, rather than the step-by-step construction described in the previous paragraph.

We can now conclude that the original logical spin system is realised in a subspace of the physical Hamiltonian,
\begin{eqnarray}
H_\mathrm{phys} = H_\mathrm{C}
+ \sum_{j = 1}^{N} h_{j} \SIGP{Z}{(0,j)} + \sum_{i = 1}^{N-1} \sum_{j = i+1}^{N} \mu_{i,j}J_{i,j} \SIGP{Z}{(i,j)}.\ \ \ \ 
\end{eqnarray}
Here $H_\mathrm{C}$ encapsulates the physics that forces the physical spins to respect the stabiliser constraints. We see that if we make the choice $\nu_{i,j}=+1$ for all $i,j$ then we recover exactly the physical prescription proposed by LHZ.

The constraint Hamiltonian can always be expressed as
\begin{eqnarray}
H_\mathrm{C} = \sum_n E_n P_n.
\end{eqnarray}
Here, $\{ E_n \}$ are eigenenergies and $\{ P_n \}$ are projectors corresponding to eigenstates of $H_\mathrm{C}$. $E_0$ is the ground energy, and $P_0$ corresponds to the ground-state subspace of $H_\mathrm{C}$.

If $H_\mathrm{C}$ satisfies the following conditions, logical spins are in the ground state of $H_\mathrm{logic}$ (the logical model) when physical spins are in the ground state of $H_\mathrm{phys}$:
\begin{itemize}
\item[(i)] $[P_0, H_\mathrm{phys}] = 0$;
\item[(ii)] $\forall [i,j]: S_{[i,j]}P_0 = \nu_{i,j}P_0$ and $\mu_{i,j} = \prod_{i'=0}^{i-1} \prod_{j'=i+1}^{j} \nu_{i',j'}$;
\item[(iii)] $\forall k,k': [P_0, \SIGL{Z}{k}] = [P_0, U_{k}\SIGL{X}{k}V_{k}] = [U_{k}, \SIGL{Z}{k'}] = [V_{k}, \SIGL{Z}{k'}] = 0$, where $U_{k}$ and $V_{k}$ are unitary operators;
\item[(iv)] and $E_0 + E_\mathrm{g} < E'_\mathrm{g}$, where $E_\mathrm{g}$ is the ground state energy of $H_\mathrm{logic}$, {$E'_\mathrm{g}$ ($E_0 + E_\mathrm{g}$) is the lowest eigenenergy of $H_\mathrm{phys}$ in the subspace $\Ide - P_0$ ($P_0$),} and $\Ide$ is the identity operator.
\end{itemize}
The proof is given in Sec.~\ref{sec:conditions}. The energy gap between the ground state and the first excited state in the physical model is $e_{\rm phys} = \min\{E'_\mathrm{g} - (E_0 + E_\mathrm{g}), e_{\rm logic}\}$, where $e_{\rm logic}$ is the energy gap in the logical model.

Obviously, one way to write down a suitable constraint Hamiltonian is simply to use the stabilisers themselves. For example if we choose $\nu_{i,j}=+1$, $\forall i,j$ which corresponds to the approach in Ref.~\cite{Lechner2015}, then we can write
\begin{eqnarray}
H_\mathrm{C} = \frac{\Delta}{2} \sum_{[i,j]} (\Ide - S_{[i,j]}).
\label{formalHC}
\end{eqnarray}
In the ground state of $H_\mathrm{C}$, $E_0 = 0$ and $S_{[i,j]} = +1$, i.e.~the ground state subspace is the subspace encoding $N$ logical spins and is $2^N$-dimensional. The projector to the ground subspace can be written as
\begin{eqnarray}
P_0 = \prod_{[i,j]} \frac{\Ide + S_{[i,j]}}{2}.
\end{eqnarray}
Since $\mu_{i,j} = +1$ ($S_{[i,j]}P_0 = P_0$) and $U_{k} = V_{k} = \Ide$, conditions (i), (ii) and (iii) are satisfied. When $\Delta$ is large enough, condition (iv) can also be satisfied.

However, a Hamiltonian formed by stabilisers involves three-body and four-body interactions, which are unphysical. The use of ancilla qutrits to achieve an equivalent but physically realistic $H_\mathrm{C}$ is discussed further in the next section.

Making other choices for the stabiliser values $\nu_{i,j}$ can lead to interesting variants. Consider, for example, the choice $\nu_{i,j}=-1$ for all $i$ and $j$. This means that each local stabiliser is requiring odd parity among its group of physical spins. Consequently some of the $\mu$ values will be $-1$ according to Eqn.~(\ref{eqn:muRule}). Specifically, $\mu_{i,j}=-1$ when $i$ is odd and $j$ is even. The three cases that occur for the $N=5$ system are marked in Fig.~\ref{fig:squareLattice}(b). Thus these particular $J_{i,j}$ couplings from the original, logical Hamiltonian are multiplied by $-1$ in the physical Hamiltonian. Presumably this will not present difficulties for any relevant hardware system, since such a system will need to handle both positive and negative $J$ values in any case in order to tackle non-trivial computation problems. But there is a more profound consequence for the hardware implementation: all our stabilisers now seek to constrain their local groups of spins to odd parities, and this may be easier to realise than the even parity constraint. Formally, a suitable $H_\mathrm{C}$ can be written in analogous terms to Eqn.~(\ref{formalHC}) as $H_\mathrm{C} = \frac{\Delta}{2} \sum_{[i,j]} (\Ide + S_{[i,j]})$, but again this is using unphysical 3 and 4 body terms. The interesting distinction is that there is now a natural way to an equivalent $H_\mathrm{C}$ using only a single ancilla qubit for each stabiliser group, as we now discuss.

\section{Ancillary-qubit Ising models}

Summarising the paper so far, the previous section introduced a stabiliser formalism and used it to map a Hamiltonian with $N$ {\it logical} spins posessing `all-to-all' interactions, to a physical Hamiltonian with $N(N+1)/2$ physical spins but requiring only local interactions. The nature of the local stabiliser rules was defined by our choice of constants $\nu_{i,j}$. We noted that the choice of setting $\nu_{i,j}=+1$ for all $i$, $j$ results in the prescription given in LHZ, i.e. the local constraints on groups of four or three spins are equivalent to demanding {\it even parity} in the number of spins aligned to the negative $z$-direction. The next most natural choice is $\nu_{i,j}=-1$ for all $i$, $j$. This leads to some $\mu_{i,j}=-1$ factors in the physical Hamiltonian, but moreover it inverts the parity requirements on all local groups from even to odd. The distinction between even and odd parity constraints seems relatively minor when the constraining Hamiltonian $H_\mathrm{C}$ is written formally using the stabilisers, as in Eqn.~(\ref{formalHC}). However, since the stabilisers are three- and four-body terms this does not suffice as a physical prescription and instead one must find a realisable $H_\mathrm{C}$ that is equivalent.

For the even parity case, LHZ suggested the introduction of an {\it ancilla} qutrit, i.e. a spin-1 system, for each group of physical spins (they remark that the role can equivalently be played with qubits rather than qutrits). Following their prescription we can write
\begin{eqnarray}
H^{\rm even}_\mathrm{C} = \frac{\Delta}{4} \sum_{[i,j]} H_{[i,j]}, \nonumber
\end{eqnarray}
where
\begin{eqnarray}
H_{[i,j]} = \left\{
\begin{array}{ll}
\left(4\TP{Z}{[i,j]}+\sum_{a}\SIGP{Z}{a}  \right)^2 & \mbox{if $i+2 < j$};  \\
\left(\Ide + 4\TP{Z}{[i,j]}+\sum_{a}\SIGP{Z}{a}  \right)^2 & \mbox{if $i+2 = j$}.
\label{LHZwithQutrit} 
\end{array} \right.
\end{eqnarray}
where $\TP{Z}{[i,j]}$ is the spin-1 (qutrit) ancilla associated with lattice face $[i,j]$, with eigenvalues $-1$, $0$ and $+1$. The sums run over the cases $(i,j)$, $(i,j-1)$, $(i+1,j)$, and $(i+1,j-1)$, or just the first three for the $i+2=j$ instances. For these latter instances, as an alternative to giving them a distinct $H_{[i,j]} $ one can instead introduce `dummy' physical spin-$\frac{1}{2}$ systems that form an additional row beneath but which are `locked' to the {$\SIGP{Z}{a} = +1$} eigenstate; the physics is identical.

The degenerate ground state of  $H^{\rm even}_\mathrm{C}$ is a subspace formed from all the `correct' even parity configurations of physical spins where each is matched with a correlated state of the ancilla spin. Note that the ancilla has no role in $H_{\rm phys}$ outside of $H^{\rm even}_\mathrm{C}$. An intuition behind the use of the ancilla is as follows: Because the term is squared, the lowest energy contribution it can make is zero. Note that the sum of the four $\SIGP{Z}{}$ operators can take values $-4$, $-2$, $0$, $2$ or $4$, and the value of the $4\TP{Z}{}$ operator can be equal to $-4$, $0$ or $4$. Therefore, if the physical spins sum to $\pm2$, there is no assignment of the qutrit that can achieve a total energy of zero; but $-4$, $0$ and $4$ are acceptable. These are of course precisely the even parity states.

Now we consider the equivalent cases for our `always odd parity' scenario, which we obtained by considering the $\nu_{i,j}=-1$ case. Again we must identify a simple physical $H_\mathrm{C}$ with a correct degenerate ground state. As with the LHZ example we again use the technique of squaring a sum of $\sum_{a}\SIGP{Z}{a} $ operators; but now we find that the ancilla need only take two values. Specifically, we opt for
\begin{eqnarray}
H^{\rm odd}_\mathrm{C} = \frac{\Delta}{4} \sum_{[i,j]} H_{[i,j]}, \nonumber
\end{eqnarray}
where
\begin{eqnarray}
H_{[i,j]} = \left\{
\begin{array}{ll}
\left(2\SIGP{Z}{[i,j]}+\sum_{a}\SIGP{Z}{a}  \right)^2 & \mbox{if $i+2 < j$;\ \ \ }\\
\left(\Ide + 2\SIGP{Z}{[i,j]}+\sum_{a}\SIGP{Z}{a}  \right)^2 & \mbox{if $i+2 = j$.\ \ \ }
\end{array} \right. \label{oddHc}
\end{eqnarray}
Here $\SIGP{Z}{[i,j]}$ (note the brackets [ ] in the subscript) is the spin-$\frac{1}{2}$ ancilla associated with lattice face $[i,j]$. As before the sums run over the cases $(i,j)$, $(i,j-1)$, $(i+1,j)$, and $(i+1,j-1)$, or just the first three for the $i+2=j$ instances. Again, one could introduce a row of `dummy' physical spins below the the active array to make all faces of the array  square so that the $i+2=j$ cases are no longer special. This follows the form of the LHZ construct exactly, except for the $2\SIGP{Z}{[i,j]}$ term which of course has values $\pm 2$. The same intuition just described therefore leads us to see that the ground state will be spanned by states where the $\sum_{a}\SIGP{Z}{a} $ yields $\pm 2$, i.e. the odd parity states. In Sec.~\ref{sec:subspace}. we show that such a constraint Hamiltonian satisfies all the conditions (i)-(iv) if $\Delta$ is large enough.

The required interactions are encouragingly simple. Assuming that the physical implementation indeed uses a row dummy spins, then expanding the squared expression in Eqn.~(\ref{oddHc}) and neglecting global shifts gives
\begin{eqnarray}
H_{[i,j]} =
\displaystyle{\sum_{a\neq b}}\SIGP{Z}{a}\SIGP{Z}{b}+2\sum_{a}\SIGP{Z}{a}\SIGP{Z}{[i,j]}
\label{eq:simpleConstraint}
\end{eqnarray}
where here the $a$ and $b$ indices run over the local physical spins as usual. This appears to be a potential advantage over the even parity solution with its qutrit (spin-1) ancilla, not only because qubit (spin-$\frac{1}{2}$) systems may be easier to realise but also because the expansion of Eqn.~(\ref{LHZwithQutrit}) has terms $(\TP{Z}{})^2$ which may be awkward to realise; for the odd-parity version, the equivalent terms $(\SIGP{Z}{[i,j]})^2$ are merely the identity, and can be neglected. 

Equation (\ref{eq:simpleConstraint}) involves interactions between all four spins defining the lattice face, and an interaction of the same form but double the strength between each of these and their shared ancilla spin. In fact the ratio of $2$ between these strengths is not required; it is the optimal, but any ratio greater than 1 but less than 3 will correctly reproduce the effect of the stabilisers. It is interesting to speculate that this set of interactions might be very naturally realised by connecting the four physical spins to a common resonator, and coupling this group's ancilla to the same resonator with twice the coupling strength. This is indicated schematically in Fig.~\ref{fig:squareLattice}(c).

{\it Note added while drafting version 2 of this preprint: Two new papers~\cite{ChancellorPreprint,LeibPreprint} discuss superconducting systems capable of supporting $M$-body parity constraints (stabilisers, in our language). A further work~\cite{ChancellorPreprint2} has shown how these $M$-body parity constraints can be implemented on the Chimera graph.}

\section{Spectrum and numerical results}

The proceeding sections provide an analytic treatment within which both the LHZ proposal, and a variant proposal based on odd parity, have emerged as examples of local Hamiltonians that can simulate all-to-all interactions. Before moving on to consider new forms of physical spin layout, we wish to compare these two alternatives using a numerical study of small systems.

Generally the analytic conclusions described earlier are valid when the energy $\Delta$ associated with the parity-constraining terms is sufficiently large compared to other terms. It is interesting to see how these two approaches perform for finite values of the parameters.

We performed our simulations using exact diagonalisation. Because the number of ancillas required is $(N-1)N/2$, the total number of physical qubits required is $(N+1)N/2 + N(N-1)/2 = N^2$ (if we consider triangular constraints in the bottom layer). This quadratic scaling severely limits our ability to simulate numerically even small systems, especially when using ancillary-qutrits (see Table~\ref{tab:size}).

\begin{table}[b]
\centering
 \begin{tabular}{||c c c||} 
 \hline
 N & qutrit & qubit \\ [0.5ex] 
 \hline\hline
  $3$ & $2^{6} 3^{3}$ & $2^{9}$ \\ 
 \hline
 $4$ & $2^{10} 3^{6}$ & $2^{16}$ \\ 
 \hline
 $5$ & $2^{15} 3^{10}$ & $2^{25}$ \\
 \hline
 $6$ & $2^{21} 3^{15}$ & $2^{36}$ \\
 \hline
 $7$ & $2^{28} 3^{21}$ & $2^{49}$ \\[0.5ex] 
 \hline
\end{tabular}
\caption{
The size of the computational space scales quadratically with the number of logical qubits $N$. Left column: size of logical system. Central column: size of the computational space for the ancillary-qutrit (even parity) architecture. Right column: size of the computational space for the ancillary qubit (odd parity) architecture.
}
\label{tab:size}
\end{table}

As discussed in \cite{Lechner2015} the strength of the constraint terms $\Delta$ is one of the key adjustable parameters of the architecture. The analytic arguments in the proceeding section rely on $\Delta$ being the dominant energy at the end of the adiabatic sweep, so that the correct stabilisers are enforced. On the other hand, we also wish for the energy scales ($h$ and $J$) to be as large as possible, since their magnitude will influence the gap between the ground state of the logical Hamiltonian and its excited states, and thus determine the speed and practicality of quantum annealing or other adiabatic processes. Consequently, it is interesting to see how close we can permit those lesser energy scales to come to $\Delta$, or in order words how modest a ratio will suffice.

Presently we note that there is another reason to be interested in modest values of this ratio, concerning the detectability and correctability of errors in the system's evolution.

In our simulations we benchmark the different parity enforcing terms with two different metrics while varying the strength of $\Delta$. We employ a random Ising Model where the $J_{ij}$ elements are drawn from a uniform distribution in $[-J, J]$. The energies $h_i$ are drawn from the same distribution. The quantity $J_{av}$ is the average unsigned value, i.e. $J/2$. We will be interested in the ratio $R$ between $\Delta$, the energy scale of the parity-constraining terms, and $J_{av}$.

\begin{figure}[b]
\centering
\includegraphics[scale=.42]{\figpath 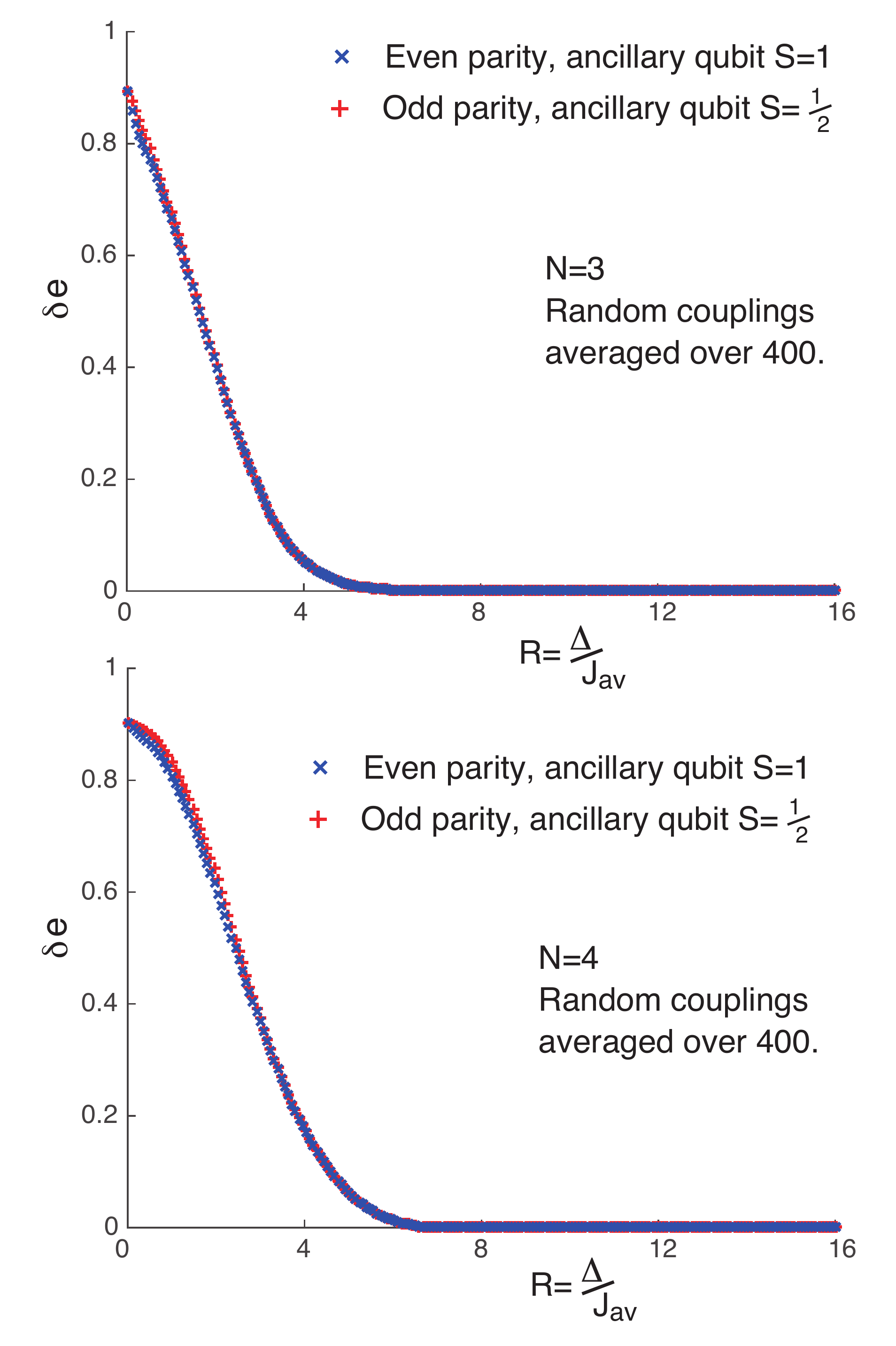}
\caption{
Deviation between the lowest energy gap in the logical Hamiltonian $H_{\rm logic}$ and the physical Hamiltonian $H_{\rm phys}$.
}
\label{fig:data_DeltaE_rand}
\end{figure}

Following along the same lines as the numerical analysis in Lechner {\it et al.}  \cite{Lechner2015}, for our first metric we take the system to be at the end of its adiabatic sweep, and we find the gap between the ground state and lowest excited state(s). We find this gap for the physical system, i.e. the $N(N+1)/2$ array of physical spins, and see how it deviates from the same quantity found using the ideal logical Hamiltonian.  That is to say, we plot
$$
\delta e = | e_{\rm logic} - e_{\rm phys} |
$$
where $e_{\rm logic} = \lambda_{\rm logic}(1) - \lambda_{\rm logic}(0) $, $e_{\rm phys} = \lambda_{\rm phys}(1) - \lambda_{\rm phys}(0) $ and $\lambda(i)$ is the $i$-th eignevalue of a given architecture. Locating the value of $\Delta$ where this deviation largely vanishes gives insight into how large $\Delta$ should be in order that the mapping process is successful. Figure~\ref{fig:data_DeltaE_rand} shows the results for systems of $N=3$ and $N=4$ logical qubits (i.e. $9$ and $16$ physical spins, respectively). The behaviour is as expected; there is no significant difference between the even parity physical architecture (with its spin-1 qutrit ancillas) and the odd parity system (using spin-$\frac{1}{2}$ qubit ancillas).

\begin{figure}
\centering
\includegraphics[scale=.42]{\figpath 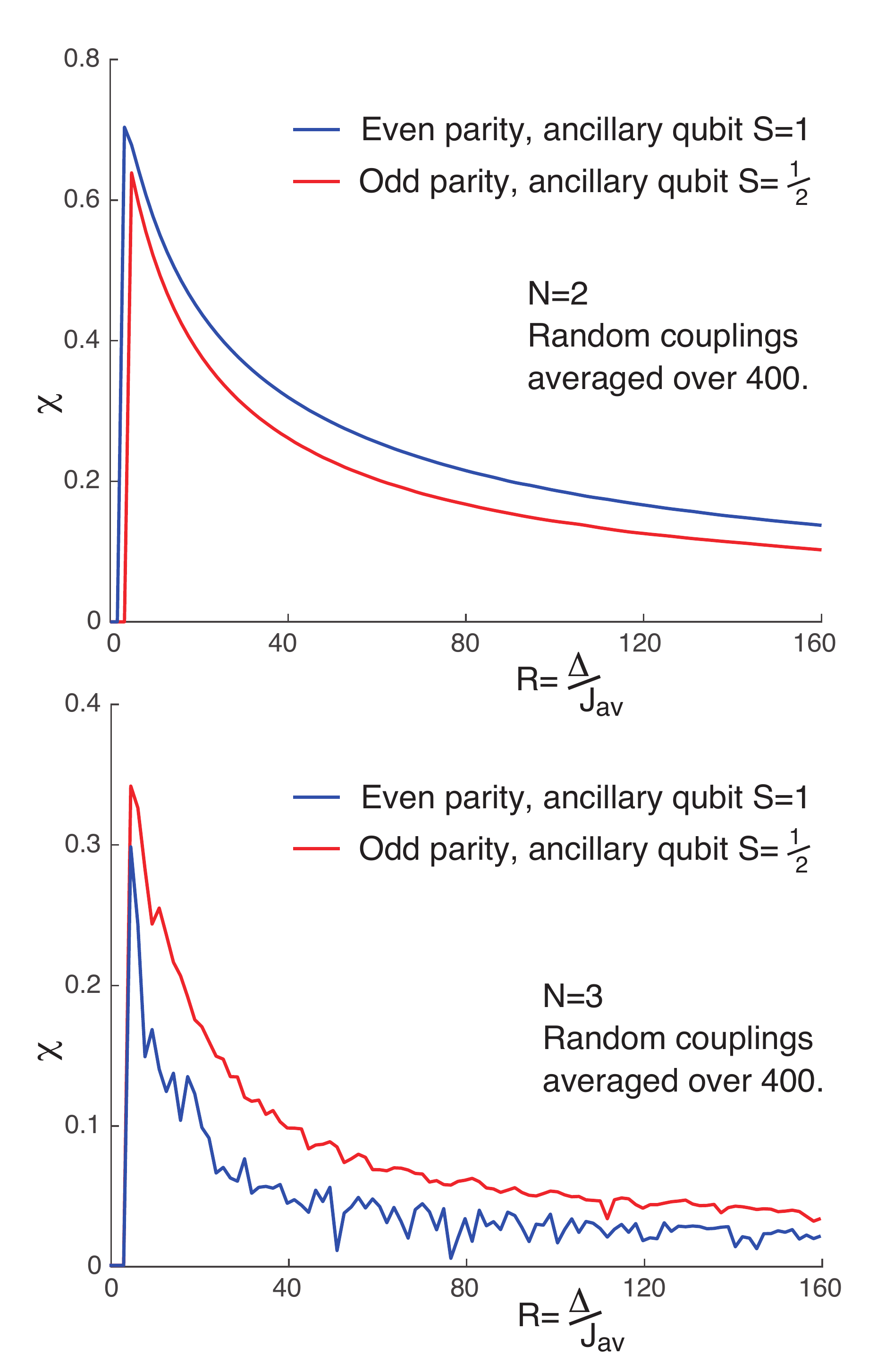}
\caption{
Minimum Gap ratio between the physical and logical systems.
}
\label{fig:data_MinGap_R}
\end{figure}

Perhaps the more crucial energy in a quantum annealing process is not the gap at the end of the anneal, but rather the minimum gap that occurs at any time throughout the evolution. It is this gap which is usually taken characterise the stability of the process and the speed with which it can be completed. Our second metric concerns this minimum gap as we sweep between an initial Hamiltonian $\sum _i \SIGP{X}{i} $ (where the sum runs over all the physical qubits) and the final form. We plot the ratio $\chi$ of the minimum gap occurring in the physical architecture (which depends on $R$) to that which would occur in the logical system:
$$
\chi(R) = \mathrm{MinGap}_{\mathrm{phys}} / \mathrm{MinGap}_{\mathrm{logic}}.
$$

In Fig.~\ref{fig:data_MinGap_R} we show the behaviour of $\chi$ for the smallest two non-trivial systems: $N=2$ and $N=3$ logical qubits. Each data point is an average of 400 simulations, and we have chosen to find the average of $\frac{1}{\chi}$ and then reciprocate; this emphasises cases where the gap in the physical system vanishes (or nearly vanishes). Interestingly we do see some variation between the behaviour of the even-parity constraining system with its spin-1 ancillas, and the alternative odd-parity architecture using spin-$\frac{1}{2}$ ancillas. The curve of the former approaches the $x$-axis many times, suggesting level crossing for the ancillary-qutrit implementation that are not present when using the ancillary-qubit version.

\section{Remarks on error detection and correction}

The analytic proportion of this paper stresses the significance of the stabiliser-enforcing terms, and may lead one to suppose that we should like to have $\Delta\gg J,h$. However from our small system numerics we note that that it can suffice for these energy ratios to be modest, and indeed from Fig.~\ref{fig:data_MinGap_R} we see that larger values of this ratio can be associated with smaller values of the gap. This may lead one to speculate that moderately large values of the ratio $R=\Delta/J_{av}$ are optimal, and indeed the following speculative line of thought leads to the same conclusion.

As pointed out by Pastawski and Preskill~\cite{PandP2015}, the multiple parity constraints applied to the physical system have the consequence that, if the final state is read out incorrectly, it is highly likely that classical post-processing can recover the correct set of measurements. The threshold for error correction, i.e. the number of spins that would need to be misread before the correct state cannot be inferred, is very high. Can this permit us to correct an error that occurs during the evolution, i.e. a jump from the ground state to an excited state? This depends on whether the excited state in question has different parity values. If $R$ is very large compared to other energies, then the spectrum will be such that the low-lying excited states have correct parity, and therefore jumps to these states are uncorrectable (indeed, undetectable except that the final measured state make constitute a poor solution to the optimisation problem!). However, opting for a more modest $R$ value may permit the low-lying excited states to violate the parity constraints, and thus conceivably permit us to correct them post-measurement.

\section{Generalised annealer codes}

\begin{figure}
\centering
\includegraphics[scale=.5]{\figpath 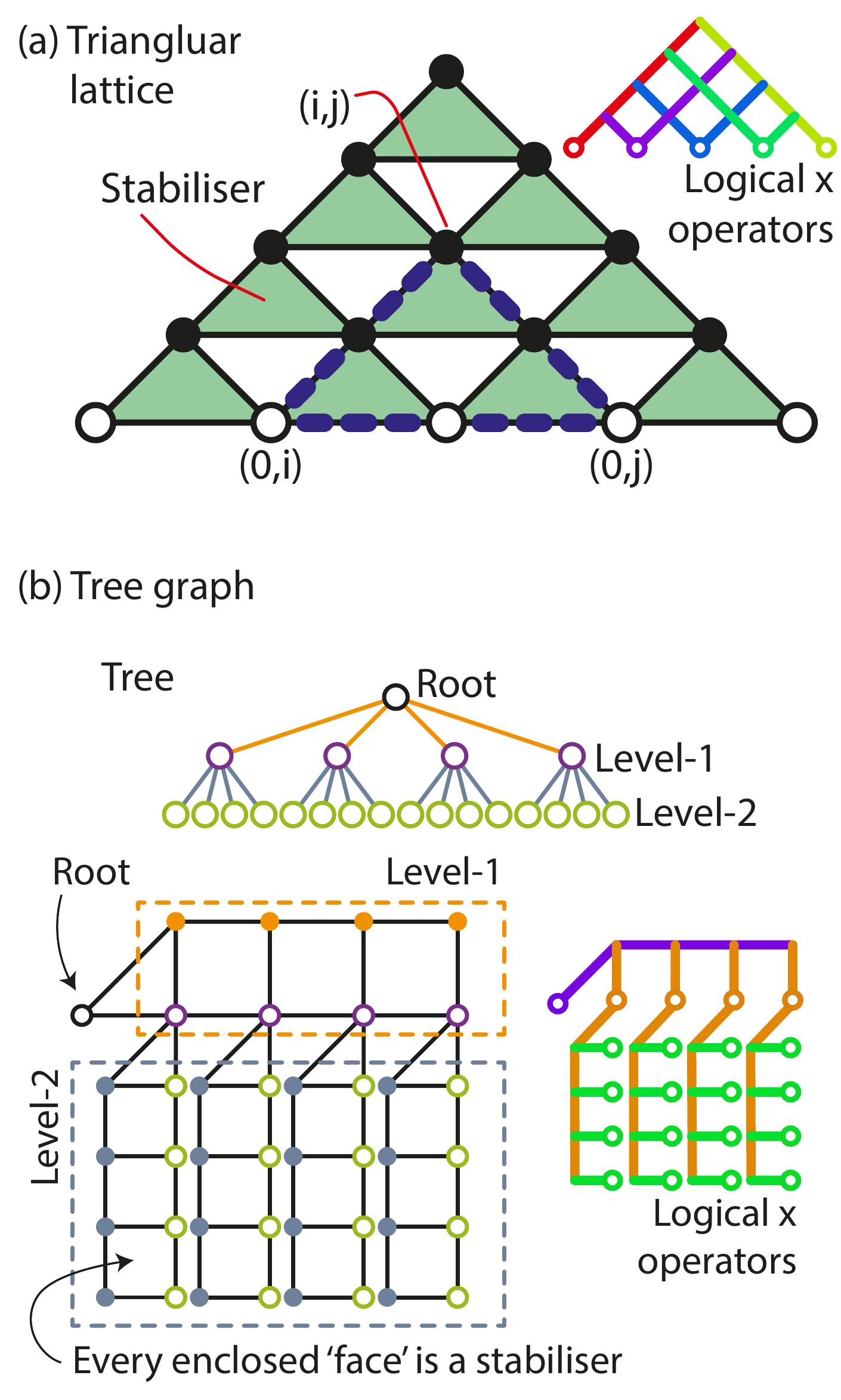}
\caption{
(a) The lattice for encoding all-to-all connected logical spins using three-spin stabilisers. Each circle denotes a physical spin. Empty circles are vertex spins $(0,j)$ and solid circles are edge spins $(i\neq 0,j)$. Spins $(0,i)$, $(0,j)$ and $(i,j)$ always form an isosceles triangle (marked with the dashed blue line). Green triangles are stabilisers, and each stabiliser has three spins. (b) Stabiliser code of the tree-graph logical model. Each empty circle denotes a vertex physical spin (or a logical spin in the tree graph), and each solid circle denotes an edge spin corresponding to the two-body Ising interaction between two logical spins. Each face corresponds to a stabiliser as in Fig.~\ref{fig:squareLattice}(a). 
}
\label{fig:triangular}
\end{figure}

So far we have used our stabiliser formalism only to recover the proposal of LHZ, with an additional freedom to choose odd versus even parity constraints. But the stabiliser picture can allow us to design a wide range of physical layouts, in order to realise different levels of connectivity and/or higher order correlations than two-body.

Before varying the nature of the logical Hamiltonian, we note that our approach can guide us to layouts which support exactly the same logical Hamintoan as the LHZ construction, but  which have different kinds of stabiliser (rather than merely different stabiliser {\it signs}, as considered earlier). Perhaps the most simple set of stabilisers is the triangular pattern shown in Fig.~\ref{fig:triangular}(a). When finding sets of stabilisers such as these, it is helpful to remember the principle that any product of stabilisers is also a stabiliser. This can allow one to translate from stabilisers that are non-local in the  physical layout, to a local set. For example, the product of the three stabilisers within the large blue dashed triangle, is equivalent to stabiliser $\SIGP{Z}{(0,i)} \SIGP{Z}{(0,j)} \SIGP{Z}{(i,j)}$ which involves only the spins at the corners of that triangle (because each spin in the middle of an edge appears twice in the product, so cancelling out). This layout, or others generated using the same principles, may prove more natural to implement with a given technology.

We now generalise away from Hamiltonians that require all-to-all connectivity, but for the moment we will continue to restrict our logical Hamiltonians to involve only one- and two-body terms. We therefore consider a connectivity graph with vertices $V$, representing the logical spins, and edges $E$ representing the required terms in the logical Hamiltonian Eqn.~(\ref{logicalH}). That is to say, if $J_{i,j}\neq 0$ then the edge linking vertex $i$ and vertex $j$ is present in set $E$. We will need $|V| + |E|$ physical spins if we wish to encode $|V| = N$ logical spins, i.e.~one physical spin $(0,i)$ for each vertex $i\in V$ and one physical spin $(i,j)$ for each edge $(i,j)\in E$. 

We need only follow our earlier prescription: We select $N$ of the physical spins to represent the logical $z$-operators, i.e. we identify the physical spins $(0,i)$ for whom $\SIGP{Z}{0,i}$ is identified with $\SIGL{Z}{i}$. We then define $|E|$ independent stabilisers, as before specifying each stabiliser as a product of $\SIGP{Z}{}$ operators. Finally we determine the logical $x$-operators that are implied by these choices, recalling the requirements: operator $\SIGL{X}{i}$ must commute with all other logical $x$-operators and with all the stabilisers, and it must commute with all $\SIGL{Z}{j\neq i}$ while anti-commuting with $\SIGL{Z}{i}$. As before, this leads us to the rule that the product of physical $\SIGP{X}{}$ operators which constitutes a given logical $x$-operator $\SIGL{X}{i}$ must (a) include $\SIGP{X}{i}$, (b) exclude all $\SIGP{X}{j\neq j}$ and (c) include an even number (or zero) of operators that address spins in each stabiliser. 

Assuming that all logical $x$-operators have been identified, we can now identify the roles of the remaining  physical spins using the following rule: The physical spin where logical $x$-operators $\SIGL{X}{i}$ and $\SIGL{X}{j}$ intersect, is to be labelled $(i,j)$. This spin's physical $z$-operator $\SIGP{Z}{i,j}$ is identical to the logical two-body term $\SIGL{Z}{i}\SIGL{Z}{j}$, up to a sign $\mu_{i,j}$. The sign is simply a function of which stabilisers we have chosen to be negative, as in the example leading to Eqn.~\ref{eqn:muRule}. In an earlier section we alluded to this convenient rule, and we now justify it.

Taken together the logical $\SIGL{Z}{}$ operators and the stabilisers form a total of $|V| + |E|$ {independent} operators, each of which is either a physical operator $\SIGP{Z}{}$ or a product of such operators. From their independence it follows that we must be able to express any operator $\SIGP{Z}{(i,j)}$ as a product of logical operators and stabilisers, i.e.~
$$
\SIGP{Z}{(i,j)} = \text{(Product of }\SIGL{Z}{}\text{)} \times \text{(Product of stabilisers)}.
$$
We can determine which logical $\SIGL{Z}{}$ operators are in this product by considering logical $\SIGL{X}{}$ operators. If $\SIGL{Z}{k}$ is in the product, $\SIGP{Z}{(i,j)}$ anti-commutes with $\SIGL{X}{k}$, otherwise $\SIGP{Z}{(i,j)}$ commutes with $\SIGL{X}{k}$. But from our definition of the logical $x$ operators, only $\SIGL{X}{i}$ and $\SIGL{X}{j}$ anti-commute with $\SIGP{Z}{(i,j)}$ so we conclude that only $\SIGL{Z}{i}$ and $\SIGL{Z}{j}$ are in the product, i.e.~
$$
\SIGP{Z}{(i,j)} = \SIGL{Z}{i}\SIGL{Z}{j} \times \text{(Product of stabilisers)}.
$$
Therefore, the Ising interaction $J_{i,j}\SIGL{Z}{i}\SIGL{Z}{j}$ in the logical model can be mapped to $J_{i,j}\SIGP{Z}{i,j}$ in the physical model  (up to a sign determined by the value of stabilisers).

\begin{figure*}
\centering
\includegraphics[scale=.37]{\figpath 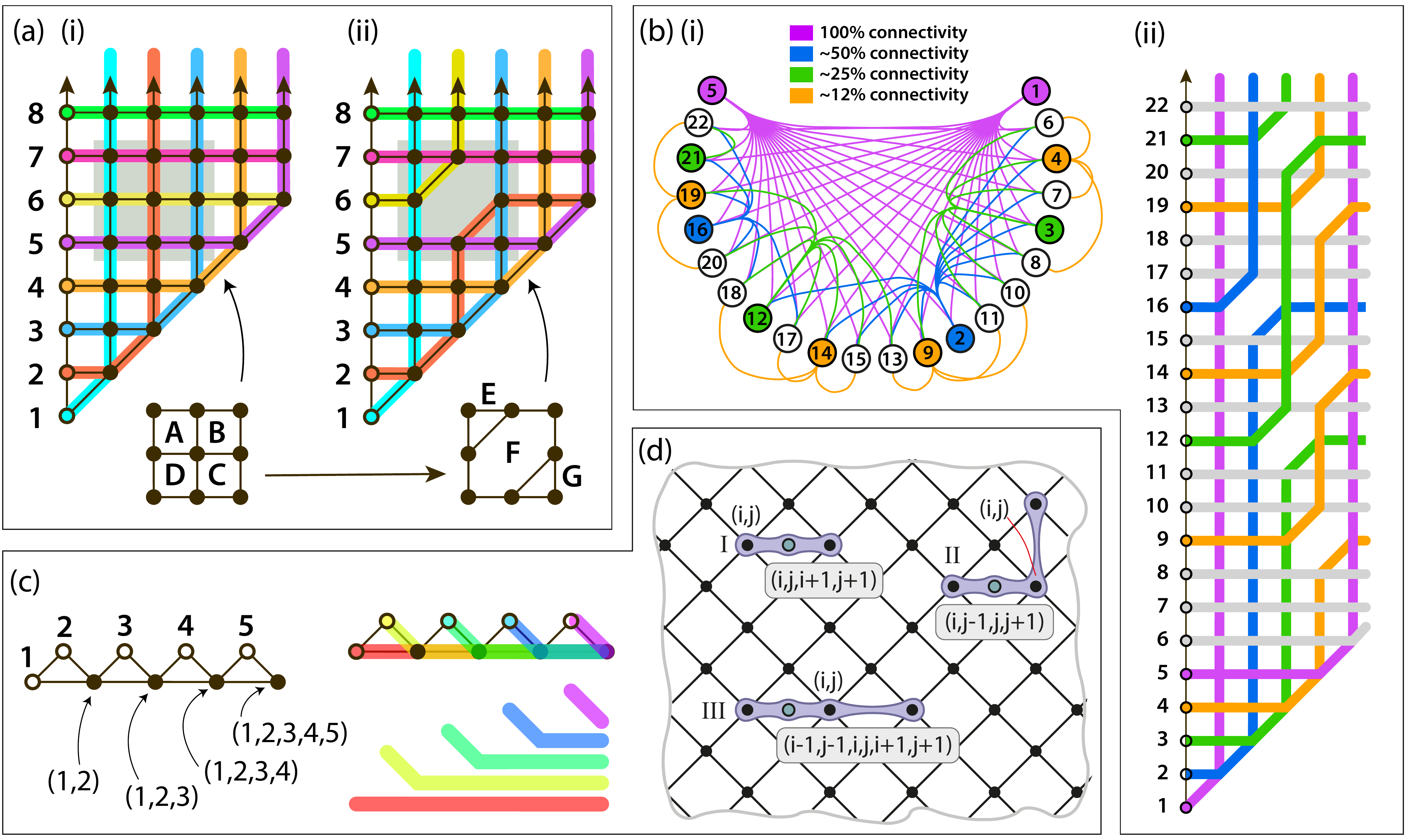}
\caption{
(a) Strategic disruptions to the stabiliser lattice can control the `routing' of logical $x$-operators. Starting from a regular lattice section (i) and replacing the central four stabilisers A-D with three stabilisers E, F, G (where F is a six-body stabiliser) and removing a physical spin, results in a new lattice (ii) where the paths of the $x$-operators for logical spins 2 and 6 are `reflected'. (b) Designing non-trivial connectivities via such reflections: Diagram (i) is a connectivity graph showing a highly connected (but not all-to-all) relationship between 22 nodes. Digram (ii) is a lattice formed from 3, 4 and 6 body stabilisers, arranged so as to realise that connectivity; for every linked pair in (i) there is at least one physical spin  in (ii) representing the relative orientation of the logical spins.
(c) and (d) show methods for representing multi-body interactions. In (c) this is achieved by a ladder of stabilisers; filled circles represent increasingly high order correlations. In (d) the lattice notation follows that of Fig.~\ref{fig:squareLattice}(a). Circles filled in green denotes additional physical spins representing multi-body interactions. Purple links denote stabilisers. In the case I, the stabiliser of physical spins $(i,j)$, $(i+1,j+1)$ and the additional physical spin corresponds to the four-body term $\SIGL{Z}{i}\SIGL{Z}{j}\SIGL{Z}{i+1}\SIGL{Z}{j+1}$; in the case II, the stabiliser of physical spins $(i-1,j-1)$, $(i,j)$, $(i-1,j+1)$ and the additional physical spin corresponds to the four-body term $\SIGL{Z}{i}\SIGL{Z}{j-1}\SIGL{Z}{j}\SIGL{Z}{j+1}$; and in the case III, the stabiliser of physical spins $(i-1,j-1)$, $(i,j)$, $(i+1,j+1)$ and the additional physical spin corresponds to the six-body term $\SIGL{Z}{i-1}\SIGL{Z}{j-1}\SIGL{Z}{i}\SIGL{Z}{j}\SIGL{Z}{i+1}\SIGL{Z}{j+1}$. 
}
\label{fig:moreExamples}
\end{figure*}

In practice this means that if we wish to have a physical spin representing a two-body term  $\SIGL{Z}{i}\SIGL{Z}{j}$ in the logical Hamiltonian, i.e. if that edge exists in $E$, then the lines of physical spins associated with the two logical $x$-operators must cross.
This provides a design principle to create a bespoke physical array to represent a given logical Hamiltonian. We can see that the square lattice of LHZ, Fig.~\ref{fig:squareLattice}, and the triangular lattice in Fig.~\ref{fig:triangular}(a), meet this condition for a fully connected graph, i.e. every logical $x$-operator intersects with every other and so all edges exist. However the layout in Fig.~\ref{fig:triangular}(b) supports a more restricted graph, i.e. a three-tier hierarchical tree, and consequently requires only $2N-1$ physical spins for $N$ logical variables. For a logical Hamiltonian with exactly this connectivity, this bespoke layout therefore provides a more efficient representation and presumably the gap during annealing may be larger.

Figure~\ref{fig:moreExamples}(a) and (b) provide further examples of interesting bespoke layouts. The layout in (a)(i) is a simple pattern with a logical connectivity graph such that vertices (i.e. logical spins) numbers $1$ to $5$ are connected to all other vertices, while vertices $6$ and higher do not interconnect among themselves. An interesting variant occurs if we remove one physical spin, and correspondingly reduce the stabiliser count by one, as shown in Fig.~\ref{fig:moreExamples}(a)(ii). Note a central group of four square stabilisers has been replaced with two triangular stabilisers and a single hexagonal stabiliser. (As an aside we note that a suitable six-body, negative-parity stabiliser can be realised using only two ancilla qubits, see Appendix~\ref{sec:ancillaConstructions}) The effect of this central disruption to the layout is to effectively `reflect' the logical $x$-operator chains that would have passed though it. This alters the logical connectivity graph, for example logical spin number $2$ now only connects to $1,3,4,5$. Interestingly there are now two physical spins for each of the labels $\SIGP{Z}{2,3}$, $\SIGP{Z}{2,4}$ and $\SIGP{Z}{2,5}$. This does not present an in-principle difficulty when translating from the logical Hamiltonian to the local fields on the physical spins, we simply need to ensure that the total field on the two spins labelled $\SIGP{Z}{2,3}$ is equal to the factor $J_{2,3}$ in the logical Hamiltonian, and similarly for the others. 

The `reflecting' stabiliser in Fig.~\ref{fig:squareLattice}(a)(ii) can be used repeatedly within a larger lattice in order to control the interactions between logical $x$-operators, and thus define the logical connectivity graph. This is illustrated in Fig.~\ref{fig:squareLattice}(c). Here the layout realises a rather complex connectivity graph in which there are a small number of highly connected vertices, a larger number of more modestly connected nodes, and so on. One might, for example, choose the width of the strip of physical qubits to be $\log(N)$ while its length is $N$, where $N$ is the number of logical spins. With these $N\log(N)$ spins one could engineer a hierarchy where one logical spin connects to all others, two connect to 50\% of the set, four connect to 25\%, and so on.

Finally we consider how the stabiliser picture presented here generalises to support terms in the logical Hamiltonian that are higher than two-body. Previously we noted that when two logical $x$-operators $i$ and $j$ intersect, the physical spin at the intersection necessarily represents the two-body logical operator $\SIGL{Z}{i}\SIGL{Z}{j}$. However, it is possible for multiple logical operators to intersect at a specific physical spin, as shown in Fig.~\ref{fig:squareLattice}(c). Then the same arguments developed above apply, so that when the logical $x$-operators for logical spins $i$, $j$, ... $p$ all intersect then physical operator $\SIGP{Z} { }$ on that spin will correspond to the logical product $\SIGL{Z}{i}\SIGL{Z}{j}...\SIGL{Z}{p}$, up to a sign determined by the use of negative stabiliser constraints.

Figure~\ref{fig:squareLattice}(d) provides a second illustration of how higher order terms can be introduced. The figures shows a region of a the standard LHZ layout, i.e. a larger version of the pattern shown in Fig.~\ref{fig:squareLattice}. Here there are three additional physical spins (green circles) and correspondingly three additional stabilisers (grey shaded regions). Each of the new physical spins provides a high order logical correlation, as specified in the caption. 

In is interesting to note that the ideas presented here could potentially be used {\it in addition} to the principle of minor embedding, rather than replacing that approach outright. Starting from the original logical Hamiltonian, which directly corresponds to the structure of some computational task, one might use the minor embedding principle to derive a second, intermediate logical Hamiltonian with a larger number of spins. This intermediate Hamiltonian could then be translated into a stabiliser based layout as described here. The potential benefit would be an increased flexibility in the connectivity offered by a given layout pattern.

\section{Conclusions}

We began by presenting a stabiliser formulation for the problem of mapping a Hamiltonian with $N$ `all-to-all' interacting spins, to a Hamiltonian of $N(N+1)/2$ spins with only local interactions. As a first illustration of the approach, we took the recent work of Lechner, Hauke and Zoller, and adopted their physical spin layout. We noted the resulting logical $z$ and logical $x$ operators; the latter are chains of operators that traverse the lattice (as occurs in topological error correcting codes). We recover the LHZ result, and we also identify an interesting variant based on constraining local groups of spins to {\it odd parity}, rather than even parity.  This variant has an attractively simple realisation in terms of pure Ising interactions and ancilla qubits (rather than qutrits), and might perhaps be realised through commonly shared resonators. We numerically verify our results for small systems of $N=2,3,4$ logical qubits. There are some indications that our new odd-parity, qubit-ancilla model may maintain a more reliable energy gap during an anneal.

Having thus demonstrated our formalism in an established context, we proceed to show how it can be used to a wide range of different physical spins layouts. We display a triangle lattice for all-to-all connectivity, before moving on to create layouts which support specific (less than all-to-all) connectivities with the advantage that fewer than Order$(N^2)$ physical spins are needed. Our examples include a three-tier tree structure requiring $2N-1$ physical spins, and a more complex pattern offering a range of connectivities with $N\log(N)$ physical spins. Finally we show that there is no constraint to two-body logical terms; even within a strictly 2D layout, arbitrarily high order logical terms can be realised in a natural way.

\begin{acknowledgments}
This work was supported by the EPSRC National Quantum Technology Hub in Networked Quantum Information Processing. AR is supported by an EPSRC DTP Scholarship. The authors would like to acknowledge the use of the University of Oxford Advanced Research Computing (ARC) facility in carrying out this work~\cite{ARC}.
\end{acknowledgments}

\appendix

\section{Sufficient condition of the parity constraint}
\label{sec:conditions}

Because $[P_0, H_\mathrm{phys}] = 0$, $P_0$ is a subspace of $H_\mathrm{phys}$, i.e.~$H_\mathrm{phys}$ can be rewritten as $H_\mathrm{phys} = P_0 H_\mathrm{phys} P_0 + (\Ide - P_0)H_\mathrm{phys}(\Ide - P_0)$. In the subspace $P_0$, the spectrum is determined by the effective Hamiltonian $H_\mathrm{eff} = P_0 H_\mathrm{phys} P_0 = H_\mathrm{phys} P_0$.

Using stabilisers, single-spin Pauli operators could be expressed as
\begin{eqnarray}
\SIGP{Z}{(i\neq 0,j)} = \SIGP{Z}{(0,i)} \SIGP{Z}{(0,j)} \prod_{i'=0}^{i-1} \prod_{j'=i+1}^{j} S_{[i',j']},
\label{eq:propagator}
\end{eqnarray}
where the product of stabilisers corresponds to a rectangular area (with a corner cut) composed by faces in the lattice (see Fig.~\ref{fig:squareLattice}): the bottom face is a triangle face rather than a square, the top-left side of the area connects spins $(0,i)$ and $(0,j)$, and the top-right side of the area connects spins $(i,0)$ and $(0,j)$. As an example, the area corresponds to $\SIGP{Z}{(i\neq 2,4)}$ is highlighted in green in Fig.~\ref{fig:squareLattice}. One can find that, in the product of stabilisers, each Pauli operator occurs for even times except $\SIGP{Z}{(0,i)}$, $\SIGP{Z}{(0,j)}$ and $\SIGP{Z}{(i,j)}$. Using Eq.~(\ref{eq:propagator}) and the condition-(ii), we have
\begin{eqnarray}
H_\mathrm{eff} = H_\mathrm{phys} P_0 = E_0 P_0 + H_\mathrm{logic} P_0.
\end{eqnarray}

$P_0$ can be written as
\begin{eqnarray}
P_0 = \sum_{\alpha} P_{0,\alpha},
\end{eqnarray}
where $\alpha = (\alpha_1,\alpha_2,\ldots,\alpha_N)$ is a string of $\alpha_k = \pm 1$, $P_{0,\alpha} = P_0 \bar{P}_{\alpha}$, and
\begin{eqnarray}
\bar{P}_{\alpha} = \prod_{k} \frac{\Ide+\alpha_{k}\SIGL{Z}{k}}{2}
\end{eqnarray}
is the projector to the subspace that logical spin operators $\{ \SIGL{Z}{k} \}$ takes eigenvalues $\{ \alpha_{k} \}$. Because $[P_0,\SIGL{Z}{k}] = 0$, $\{ P_{0,\alpha} \}$ are projectors, i.e.~$P_{0,\alpha}^2 = P_{0,\alpha}$. 

For any two sets of eigenvalues $\alpha$ and $\alpha'$, we introduce a unitary operator
\begin{eqnarray}
U_{\alpha,\alpha'} = \prod_{k} [ \delta_{\alpha_{k},\alpha_{k}'}\Ide + (1-\delta_{\alpha_{k},\alpha_{k}'}) U_{k}\SIGL{X}{k}V_{k} ].
\end{eqnarray}
Then,
\begin{eqnarray}
U_{\alpha,\alpha'} P_{0,\alpha'} U_{\alpha,\alpha'}^{\dag} = P_{0,\alpha}.
\end{eqnarray}
Here, we have used the condition-(iii). Therefore, dimensions of subspaces $\{P_{0,\alpha}\}$ are the same, and the dimension $D = \Tr(P_{0,\alpha}) = \Tr(P_0)/2^N$.

Common eigenstates of $\{\SIGL{Z}{k}\}$ are eigenstates of $H_\mathrm{logic}$, and the eigenvalue only depends on $\alpha$, i.e.
\begin{eqnarray}
H_\mathrm{logic} \bar{P}_{\alpha} = \bar{E}_{\alpha} \bar{P}_{\alpha},
\end{eqnarray}
where
\begin{eqnarray}
\bar{E}_{\alpha} = \sum_{i = 1}^{N} h_{i} \alpha_{i} + \sum_{i = 1}^{N-1} \sum_{j = i+1}^{N} J_{i,j} \alpha_{i}\alpha_{j}.
\end{eqnarray}
Then, the effective Hamiltonian could be rewritten as
\begin{eqnarray}
H_\mathrm{eff} = \sum_{\alpha} (E_0 + \bar{E}_{\alpha}) P_{0,\alpha}.
\end{eqnarray}
Therefore, in the subspace $P_0$, $H_\mathrm{eff}$ (i.e.~$H_\mathrm{phys}$) and $H_\mathrm{logic}$ have the same spectrum, and dimensions of eigenenergy subspaces of $H_\mathrm{eff}$ are increased by a factor $D$.

The ground state energy of $H_\mathrm{eff}$ is $E_{0,\mathrm{g}} = E_0 + \min\{\bar{E}_{\alpha}\} = E_0 + E_\mathrm{g}$. Assuming $\alpha_\mathrm{g}$ corresponds to the ground state, then the ground state of $H_\mathrm{logic}$ is $\bar{P}_{\alpha_\mathrm{g}}$, and the ground state subspace of $H_\mathrm{eff}$ is $P_{0,\alpha_\mathrm{g}}$, i.e.~in the ground state subspace of $H_\mathrm{eff}$, logical spins are in the ground state of $H_\mathrm{logic}$.

The whole spectrum of $H_\mathrm{phys}$ is composed by the spectrum of $H_\mathrm{eff}$ and the spectrum of $(\Ide - P_0)H_\mathrm{phys}(\Ide - P_0)$. When $E_{0,\mathrm{g}} < E'_\mathrm{g}$ ($E_0 + E_\mathrm{g} < E'_\mathrm{g}$), $E_{0,\mathrm{g}}$ is the ground state energy, and $P_{0,\alpha_\mathrm{g}}$ is the ground state subspace of the Hamiltonian $H_\mathrm{phys}$.

\section{Group subspace of the ancillary-qubit model}
\label{sec:subspace}

We define $M_{[i,j]}$ as the number of excitations (number of spins along the $-z$ direction), and
\begin{eqnarray}
M_{[i,j]} = \left\{
\begin{array}{ll}
2\Ide - \frac{1}{2}\sum_{a}\SIGP{Z}{a}, & \mbox{if $i+2 < j$};\\
\frac{3}{2}\Ide - \frac{1}{2}\sum_{a}\SIGP{Z}{a}, & \mbox{if $i+2 = j$}.
\end{array} \right.
\end{eqnarray}
Here, the sums run over the cases $(i,j)$, $(i,j-1)$, $(i+1,j)$, and $(i+1,j-1)$, or just the first three for the $i+2=j$ instances. Then, in the ground state of $H_\mathrm{C}$, $E_0 = 0$, $M_{[i,j]} = 1$~or~$3$, and $S_{[i,j]} = -1$ for all stabilisers.

The projector to the ground state subspace is 
\begin{eqnarray}
P_0 = \prod_{[i,j]} P_{[i,j]},
\end{eqnarray}
where
\begin{eqnarray}
P_{[i,j]} = P_{[i,j]}^{(1)}\frac{\Ide-\SIGP{Z}{[i,j]}}{2} + P_{[i,j]}^{(3)}\frac{\Ide+\SIGP{Z}{[i,j]}}{2},
\end{eqnarray}
$P_{[i,j]}^{(m)}$ is the projector to the subspace with $M_{[i,j]} = m$, and
\begin{eqnarray}
P_{[i,j]}^{(m)} = f_m^{-1} \prod_{n=0}^{M_\mathrm{max}} [M_{[i,j]} - (n-\delta_{m,n}) \Ide].
\end{eqnarray}
Here, $M_\mathrm{max}$ is the maximum number of excitations, i.e.
\begin{eqnarray}
M_\mathrm{max} = \left\{
\begin{array}{ll}
4, & \mbox{if $i+2 < j$};\\
3, & \mbox{if $i+2 = j$}.
\end{array} \right.
\end{eqnarray}
and
\begin{eqnarray}
f_m = \prod_{n=0}^{M_\mathrm{max}} [m - (n-\delta_{m,n})].
\end{eqnarray}
Therefore, $[P_0, H_\mathrm{phys}] = 0$ and $S_{[i,j]}P_0 = - P_0$. Taking $\mu_{i,j} = (-1)^{i(j-i)}$, conditions (i) and (ii) are satisfied.

We take
\begin{eqnarray}
U_{k}^\dag = V_{k} = \prod_{[i,j]} (\sum_{m\neq 1} P_{[i,j]}^{(m)} + P_{[i,j]}^{(1)} \SIGP{X}{[i,j]}),
\end{eqnarray}
which are unitary operators describing controlled flip operations on ancillary spins. The ancillary spin $[i,j]$ is flipped iff $M_{[i,j]} = 1$. Then, we have
\begin{eqnarray}
U_{k} P_{[i,j]} U_{k}^\dag = \frac{\Ide - S_{[i,j]}}{2} \frac{\Ide + \SIGP{Z}{[i,j]}}{2},
\end{eqnarray}
i.e.~$P_0$ is $2^N$-dimensional.
Because
\begin{eqnarray}
[\frac{\Ide - S_{[i,j]}}{2} \frac{\Ide + \SIGP{Z}{[i,j]}}{2}, \SIGL{X}{k}] = 0,
\end{eqnarray}
we have $[P_{[i,j]}, U_{k}\SIGL{X}{k}V_{k}] = 0$. Therefore, $[P_0, U_{k}\SIGL{X}{k}V_{k}] = 0$, and the condition-(iii) is satisfied.

When $\Delta$ is large enough, condition (iv) can be satisfied.

\section{Ancilla constructions for many-body stabilisers}
\label{sec:ancillaConstructions}

Suppose that we wish to constrain some odd number $M$ of physical spins to a given eigenvalue, $+1$ or $-1$, of the stabiliser $\SIGP{Z}{1}\SIGP{Z}{2}\dots\SIGP{Z}{M}$. Then we can do so by including the following term into the constraint Hamiltonian $H_C$,
\[
\left(  \sum_{i=1}^M \SIGP{Z}{i} \mp \left(\mathbf{I}+2\sum_{j=1}^{P}\SIGP{Z}{[j]}\right)\right)^2.
\]
Here, as in the main text, the square brackets $[\ ]$ in the subscript denote an ancilla, and the number $P$ of ancillas is $(M-1)/2$. The term in the inner brackets has eigenvalues $M,M-4,...,2-M$. These are precisely the permitted values of the sum $\sum_{i=1}^M \SIGP{Z}{i}$ if we are in a positive eigenstate of the stabiliser, so that subtracting them implies that only the acceptable states can achieve the minimum energy of this complete term (the minimum being zero, since it is squared). The negative stabiliser is enforced by choosing to add rather than subtract the inner bracket, for an analogous reason.

Multiplying out this term will produce $ \SIGP{Z}{} \SIGP{Z}{}$ terms between the various spins involved, as well as a series of single spin terms which must be accounted for (in the case of the physical spins) by adjusting the on-site $J$ values.

Notice that for this case of odd $M$, the same number of ancilla spins are required regardless of whether we wish to enforce a positive or a negative value for the stabiliser. If instead we wish to constrain some even number $M$ of physical spins to a given eigenvalue of our stabiliser, then the number of ancillas required depends on the chosen sign. In either case, the form of the term which we should include in $H_C$ is the following:
\[
\left(  \sum_{i=1}^M \SIGP{Z}{i} +2\sum_{j=1}^{P}\SIGP{Z}{[j]}\right)^2
\]
However the number of ancillas $P$ is equal to $M/2$ if the stabiliser constraint is negative, and $M/2+1$ if the constraint is positive. This is because the positive stabiliser eigenstates have $M-1$ different possible eigenvalues of the total $z$-spin, $ \sum_{i=1}^M \SIGP{Z}{i}$, (namely, $M,M-4,..,-M$). Meanwhile the negative stabiliser eigenstates have $M-2$ possible eigenvalues of the total $z$-spin (namely $M-1,M-5,...,1-M$).

For the case of $M=4$ spin stabilisers, following LHZ one can use a qutrit rather than a qubit. But regardless this choice, the ancilla structure is more simple if one opts to enforce negative stabilisers.

\end{document}